\documentclass{aastex631}

\begin{document}

\title{Super-SNID : an expanded set of SNID classes and templates for the new era of wide-field surveys}

\author[0009-0000-6521-8842]{Dylan Magill}
\affiliation{Astrophysics Research Centre, School of Mathematics and Physics, Queens University Belfast, Belfast BT7 1NN, UK}

\author[0000-0003-1916-0664]{Michael D. Fulton}
\affiliation{Astrophysics Research Centre, School of Mathematics and Physics, Queens University Belfast, Belfast BT7 1NN, UK}

\author[0000-0002-2555-3192]{Matt Nicholl}
\affiliation{Astrophysics Research Centre, School of Mathematics and Physics, Queens University Belfast, Belfast BT7 1NN, UK}

\author[0000-0002-8229-1731]{Stephen J. Smartt}
\affiliation{Astrophysics sub-Department, Department of Physics, University of Oxford, Keble Road, Oxford, OX1 3RH, UK}
\affiliation{Astrophysics Research Centre, School of Mathematics and Physics, Queens University Belfast, Belfast BT7 1NN, UK}

\author[0000-0002-4269-7999]{Charlotte R. Angus}
\affiliation{Astrophysics Research Centre, School of Mathematics and Physics, Queens University Belfast, Belfast BT7 1NN, UK}

\author[0000-0003-4524-6883]{Shubham Srivastav}
\affiliation{Astrophysics sub-Department, Department of Physics, University of Oxford, Keble Road, Oxford, OX1 3RH, UK}

\author[0000-0001-9535-3199]{Ken W. Smith}
\affiliation{Astrophysics sub-Department, Department of Physics, University of Oxford, Keble Road, Oxford, OX1 3RH, UK}
\affiliation{Astrophysics Research Centre, School of Mathematics and Physics, Queens University Belfast, Belfast BT7 1NN, UK}



\section{Abstract} \label{sec:abstract}

We present an expanded template library for the supernova identification (SNID) software, along with updated source files that make it easy to merge our templates—and other major SNID libraries—into the base code. This expansion, dubbed 'Super-SNID', increases the number of spectra for under-represented supernova classes (e.g., SNe Ia-02cx, Ibn) and adds new classes (e.g., SLSNe, TDEs, LFBOTs). Super-SNID includes 841 spectral templates for 161 objects, primarily from the Public ESO Spectroscopic Survey of Transient Objects (PESSTO) Data Releases 1–4. The library is available on GitHub with simple installation instructions.

\section{Introduction} \label{sec:intro}

SNID (Supernova Identification) \citep{Blondin2007} is a widely used tool for determining the type, redshift and age from explosion of a supernova (SN) spectrum via template cross-matching. However, many newly discovered or rare transient classes remain unrepresented in its current library.

A surge in transient detections is expected in the coming years, driven by the Vera C. Rubin Observatory's 10-Year Legacy Survey of Space and Time \citep[LSST;][]{LSST}. Spectra for many events will be obtained by projects such as Son-of-X-shooter \citep[SOXS;][]{SOXS} and the Time-Domain Extragalactic Survey \citep[TiDES;][]{TiDES}. However, reliable classification depends on representative comparison libraries. Classes like superluminous supernovae (SLSNe) and tidal disruption events (TDEs) are not included within the standard SNID library -- as they were unknown at that time -- but are now regularly detected.

We present Super-SNID, an expanded SNID template library including spectra of newly discovered and rare transients, integrated with existing SNID libraries \citep{Blondin2007,Berkeley2012,Modjaz2016}. It is available on GitHub\footnote{\url{https://github.com/dkjmagill/QUB-SNID-Templates}; Zenodo dataset -\dataset[Super-SNID v1.0.0]{10.5281/zenodo.15167198}
} along with installation instructions, updated SNID source files to incorporate new transient classes, and a 'Template Generator' tool for adding additional spectra.

\section{Method and Results} \label{sec:rationale}

Our expanded SNID library contains several new classes, which required modifying SNID's internal source files. New classes for which we provide templates are: SLSNe\footnote{SNID templates have been created for SLSNe by \citet{Liu2016}, but the class does not exist within SNID's class definition files.}, TDEs, kilonovae, luminous fast blue optical transients (LFBOTs), and `Gap' transients (including luminous red novae and luminous blue variable eruptions). We also add several sub-classes to the existing classes of Type Ia and core-collapse SNe. The full breakdown of classes and sub-classes is given in Table \ref{tab:breakdown}. We have also added to the SNID source files several classes listed on the Transient Name Server\footnote{\url{https://www.wis-tns.org/}} for which we do not currently have templates; these are listed in the Table caption. Including these in SNID's class definitions now will simplify the future addition of templates.

We have created 841 new SNID spectral templates for 161 new objects, mainly using Data Releases 1-4 of the Public ESO Spectroscopic Survey of Transient Objects \citep[PESSTO and ePESSTO;][]{PESSTO}. 
Each spectrum was downloaded from the Weizmann Interactive Supernova Data Repository \citep[WISeREP;][]{WISeREP}. We downloaded further data for some well-studied literature objects that were not covered by PESSTO alone. Spectra of the kilonova AT2017gfo were sourced from the ENGRAVE data release \citep{2017Smartt, 2017Pian}. Additional spectra of the LFBOTs AT2018cow \citep{2018Prentice}, CSS1601010 \citep{2024Guitierrez} and AT2020xnd \citep{2021Perley} were downloaded from WISeREP. Spectra of SN Icn SN2021csp were sourced from \citet{Perley2022, Pellegrino2022}. Spectra for various SNe Ia-02es were sourced from \citet{Ganeshalingam2012,Blondin2012,Foley2010,Yi2015,Xi2024,Maguire2022}.

Several previous SNID template libraries exist, including the standard templates \citep{Blondin2007} and the additional libraries from the Berkeley Ia Supernova Program \citep{Berkeley2012}, the NYU Group stripped-envelope SN templates \citep{Liu2016, Modjaz2016, Williamson2019}, and the Type II SN templates from \citet{Gutierrez2017}. These templates are available from the SNID webpage\footnote{\url{https://people.lam.fr/blondin.stephane/software/snid/}}. To make the largest template database accessible to any SNID user, we have merged the standard SNID, Berkeley and NYU templates with our own templates in a single directory. We did not include the \citet{Gutierrez2017} templates for SNe II, as this would require maintaining an additional list of explosion dates, but this class is well covered by PESSTO SSDR1-4. By downloading our merged template directory and updated SNID source files from GitHub, a user can immediately recompile SNID to work with all templates. We hope this will serve as a useful resource to the transient community as we move towards the data-driven era of LSST, SoXS and TiDES.
We emphasise that users of Super-SNID must cite the original source papers for existing templates used, in addition to this work.

\begin{table}[]
    \centering
    \begin{tabular}{c|c|c|c|c|c}
        Type & Sub-type & No. in Standard SNID & No. with Berkeley/NYU & No. in Super-SNID & References  \\
        \hline
        Ia & Ia-norm & 1995 & 2050 & 2050 & 1, 3 \\ 
        Ia & Ia-91T & 340 & 349 & 388 & 1, 2, 3  \\
        Ia & Ia-91bg & 193 & 196 & 218 & 1, 2, 3  \\
        Ia & Ia-csm & 27 & 27 & 38 &  1, 2  \\
        Ia & Ia-pec & 103 & 103 & 138 & 1, 2  \\
        Ia & Ia-02cx & 10 & 11 & 18 & 1, 2, 3  \\
        Ia & Ia-03fg & 56 & 56 & 56 & 1  \\
        Ia & Ia-02es & 0 & 0 & 16 & 4, 5, 6, 7, 8 \\
        Ia & Ia-Ca-rich & 0 & 0 & 9 & 2  \\
        Ib & Ib-norm & 78 & 365 & 371 & 1, 2, 3, 9, 10, 11 \\
        Ib & Ib-pec & 0 & 10 & 15 & 2, 9, 10, 11 \\
        Ib & IIb & 91 & 289 & 338 & 1, 2, 3, 9, 10, 11 \\
        Ib & Ibn & 0 & 36 & 52 & 2, 9, 10, 11 \\
        Ic & Ic-norm & 82 & 294 & 341 & 1, 2, 3, 9, 10, 11 \\
        Ic & Ic-Broad & 0 & 214 & 235 & 2, 9, 10, 11 \\
        Ic & Icn & 0 & 0 & 2 & 12, 13  \\
        Ic & Ic-Ca-rich & 0 & 0 & 4 & 2  \\
        II & IIP & 190 & 339 & 601 & 1, 2, 3  \\
        II & II-pec & 241 & 241 & 247 & 1, 2  \\
        II & IIn & 75 & 143 & 205 & 1, 2, 3  \\
        II & IIL & 27 & 27 & 38 & 1, 2  \\
        II & IIn-pec & 0 & 0 & 6 & 2  \\
        NotSN & AGN & 1 & 1 & 1 & 1  \\
        NotSN & Gal & 11 & 11 & 11 & 1  \\
        NotSN & QSO & 0 & 4 & 4 & 3  \\
        NotSN & M-star & 11 & 11 & 11 & 1  \\
        NotSN & C-star & 0 & 3 & 3 & 3  \\
        SLSN & SLSN-I & 0 & 141* & 203 & 2, 9, 10, 11 \\
        SLSN & SLSN-II & 0 & 3* & 12 & 2, 3  \\
        SLSN & SLSN-IIn & 0 & 11* & 11 & 3  \\
        LFBOT & 18cow & 0 & 0 & 11 & 14  \\
        LFBOT & 20xnd & 0 & 0 & 10 & 15, 16  \\
        TDE & TDE-H & 0 & 0 & 14 & 2  \\
        TDE & TDE-He & 0 & 0 & 6 & 2  \\
        TDE & TDE-H-He & 0 & 0 & 36 & 2  \\
        TDE & TDE-Ftless & 0 & 0 & 10 & 2  \\
        KN & 17gfo & 0 & 0 & 10 & 17, 18 \\
        Gap & LRN & 0 & 0 & 3 & 2 \\
        Gap & LBV & 15 & 15 & 18 & 1, 2  \\
    \hline
    \end{tabular}
    \caption{Populated spectral classes within our updated SNID library. Columns show the number of spectra in standard SNID, after merging the Berkeley/NYU libraries, and the total in Super-SNID after merging our new templates. These numbers are accurate as of April 2025. We have added further subtypes to the source files, not shown in the table as they are currently unrepresented by any spectra. These are: Ib-Ca-rich, Ib-csm, Ic-pec, Ic-csm, Afterglow, Nova, CV, SLSN-Ib, SLSN-Ic and ILRT. 
1 = \cite{Blondin2007},  
2 = \cite{PESSTO},  
3 = \cite{Berkeley2012},  
4 = \cite{Ganeshalingam2012},  
5 = \cite{Blondin2012},  
6 = \cite{Foley2010},  
7 = \cite{Xi2024},  
8 = \cite{Maguire2022},  
9 = \cite{Liu2016},  
10 = \cite{Modjaz2016},  
11 = \cite{Williamson2019},  
12 = \cite{Perley2022},  
13 = \cite{Pellegrino2022},  
14 = \cite{2018Prentice},  
15 = \cite{2021Perley},  
16 = \cite{2024Guitierrez},  
17 = \cite{2017Smartt},  
18 = \cite{2017Pian}.  
*Class not included in standard SNID files. Adjustments made to metadata files for inclusion within Super-SNID.}
    \label{tab:breakdown}
\end{table}

\begin{acknowledgments}
DM acknowledges a studentship funded by the Leverhulme Interdisciplinary Network on Algorithmic Solutions.
MN and CRA ERC Starting Grant No.~948381. 
SJS, KS and SS acknowledge STFC Grant ST/Y001605/1, a Royal Society Research Professorship and the Hintze Charitable Foundation.
Based on observations made with ESO Telescopes	under programmes 188.D-3003 and 191.D-0935 (PESSTO/ePESSTO) and 1102.D-0353, 0102.D-0348, 0102.D-0350 (ENGRAVE). 
\end{acknowledgments}

%

\vspace{5mm}
\facilities{NTT, VLT, TNS}


\software{Astropy \citep{Astropy},  
          Numpy \citep{Numpy},
          SNID \citep{Blondin2007},
          Pandas \citep{Pandas}
          }



\vspace{5mm}


\bibliography{sample631}{}
\bibliographystyle{aasjournal}



\end{document}